\documentclass[onecolumn,sort&compress,numbers]{els-mrw} 

\usepackage{amsmath,amssymb,amsfonts,amsthm,makeidx,graphicx,mathrsfs}
\usepackage{txfonts}
\usepackage{helvet}
\usepackage{bm,bbm}

\newcommand{\eq}{\begin{eqnarray}}
\newcommand{\en}{\end{eqnarray}}
\newcommand {\lrarrow}{{\stackrel{\leftrightarrow}{\nabla}}}
\newcommand {\lrarrowD}{{\stackrel{\leftrightarrow}{\bm{D}}}}
\begin{document}


\chapter{Hadronic atoms}\label{chap1}

\author[1,2]{Akaki Rusetsky}%

\address[1]{\orgname{University of Bonn},
  \orgdiv{Helmholtz-Institut f\"ur Strahlen- und Kernphysik and\\\,
    Bethe Center for Theoretical Physics}, \orgaddress{53115 Bonn, Germany}}
\address[2]{\orgname{Tbilisi State University}, \orgdiv{High Energy Physics Institute}, \orgaddress{0186 Tbilisi, Georgia}}

\articletag{Chapter Article tagline: update of previous edition, reprint.}

\maketitle

\begin{abstract}[Abstract]
  We give a brief survey of the theory of hadronic atoms, which represent
  important sources of information for studying hadron interactions at very low
  energy. It will be namely demonstrated that a systematic expansion of the
  observables
  of hadronic atoms (the energy levels and the decay width) in terms of the
  fine-structure constant $\alpha$ can be obtained, using the framework of
  non-relativistic effective Lagrangians. We also present a pedagogical
  introduction to the non-relativistic effective theories that includes a review of the
  main concepts, such as the scale separation, construction of the Lagrangian, power
  counting and matching.

\end{abstract}

\begin{keywords}
  	Hadronic atom\sep Scattering length\sep Chiral Perturbation Theory \sep
        Non-relativistic effective theory \sep Matching
\end{keywords}

\begin{figure}[h]
	
  \hspace*{1.cm}\includegraphics[width=7.8cm]{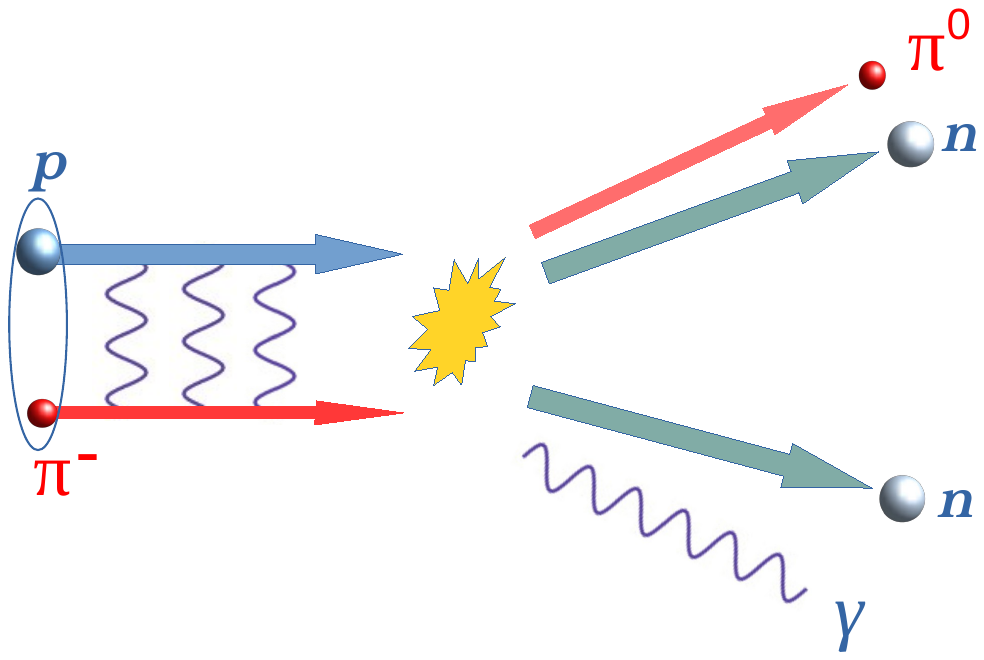}
  
  \vspace*{-10.2cm}
  
  \hspace*{10.5cm}\includegraphics[width=3.8cm]{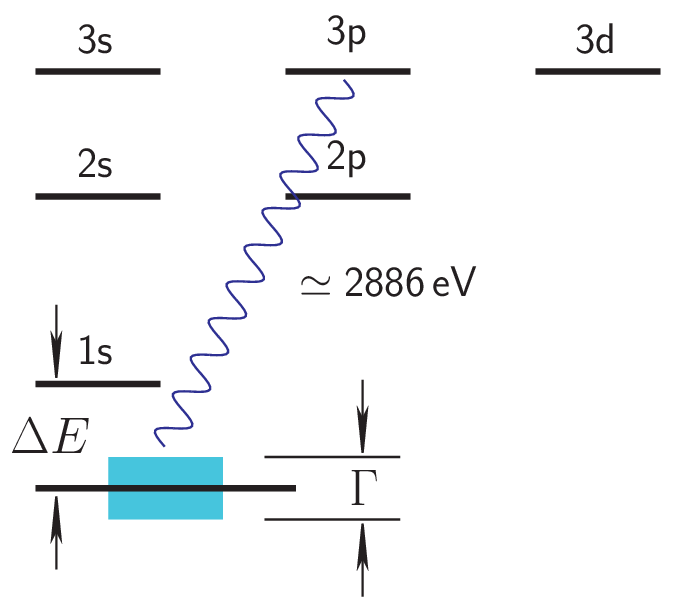}
  
  \begin{center}
    
    \caption{A schematic picture of pionic hydrogen: the average
          distance between the pion and the proton in the atom is much larger than the effective radius of strong interactions and, therefore, the strong as well as residual electromagnetic interactions give only small corrections $\Delta E$ to the Coulomb energy levels. The atom decays into different channels: strong $\pi^-p\to\pi^0n$, electromagnetic $\pi^-p\to n\gamma$ and weak
          $\pi^-\to\mu^-\bar\nu_\mu$. The contribution of the two former processes to the
          $3p-1s$ transition line width $\Gamma$ is comparable, whereas the latter (not shown) gives tiny
          contribution that can be safely neglected.}
	\label{fig:titlepage}
\end{center}        
\end{figure}

\begin{glossary}[Nomenclature]
	\begin{tabular}{@{}lp{34pc}@{}}
	  QCD & Quantum Chromodynamics\\
	  QED & Quantum Electrodynamics\\
          ChPT & Chiral Perturbation Theory\\
	  EFT & Effective field theory\\
          NREFT & Non-relativistic effective field theory\\
          CM & Center-of-mass \\
          NRQCD & Non-relativistic QCD
	\end{tabular}
\end{glossary}

\section*{Objectives}
\begin{itemize}
\item Determination of the hadron-hadron scattering lengths
  from the experiments on hadronic atoms.
\item Splitting of the strong and electromagnetic interactions; A consistent definition
  of ``purely hadronic'' quantities.
\item
  Essentials of the non-relativistic effective theory: scale separation, symmetries,
  construction of  the Lagrangian, power counting, threshold expansion, matching.
\item
  Description of the bound states.
  
\end{itemize}

\section{Introduction}\label{intro}

Imagine a hydrogen atom where the electron is substituted by a negatively charged pion,
see Fig.~\ref{fig:titlepage}.
Neglecting first all strong, weak and electromagnetic interactions other than the static
Coulomb force between two charged particles, it is seen that the spectrum of the bound
system will be given by the well-known non-relativistic formula from the textbooks on quantum mechanics:
\eq\label{eq:hydrogen}
E_n=m_p+M_\pi-\frac{\mu_c\alpha^2}{2n^2}\, ,\quad\quad n=1,2,\cdots\, .
\en
Here, $m_p$ and $M_\pi$ denote the masses of the proton and the charged pion,
respectively, and $\alpha\simeq 1/137$ is the electromagnetic fine-structure constant.
Furthermore, the reduced mass $\mu_c$ and the Bohr radius $r_B$ are given by
\eq
\mu_c=\frac{m_pM_\pi}{m_p+M_\pi}\, ,\quad\quad 
r_B=\frac{1}{\alpha\mu_c}\simeq 220\,\mbox{fm}\, .
\en
Hence, an average distance between the constituents, which is determined by the Bohr
radius, is much larger than a typical radius of strong interactions (a few fermi).
The (short-range) strong interactions practically vanish
at such distances and contribute little to the spectrum which is mainly described by
a static Coulomb force acting between the proton and the charged pion. This small deviation, however, can be measured in the experiment very accurately and provides
a unique possibility to determine the low-energy characteristics of hadron-hadron
interactions at a precision that is difficult to achieve by using other experimental methods.
Indeed, a huge difference between the Bohr radius and the effective radius of strong interactions leads to the fact that the wave function of a hadronic atom is essentially non-zero
only at a very small momenta (at the hadronic scale), i.e., the energy shift is sensitive
exclusively to the contributions from the vicinity of the elastic threshold that are given
in terms of the pion-nucleon scattering lengths.

A fundamental difference of the pionic hydrogen from the ordinary one is that the former
decays whereas the ground state of the latter is stable. The decay proceeds via different
mechanisms. Namely, the proton and the pion in the atom can come close to each other
and annihilate either into $\pi^0 n$ or $n\gamma$ final state. Albeit these
two decays are governed by different dynamics (strong and electromagnetic), the decay probability in the two channels is comparable in size, owing to the
fact that the available phase space in the second case is much larger. The decay rate
into other channels is much smaller and can be neglected. Note also that, according
to Eq.~(\ref{eq:hydrogen}), the binding energy for the ground state is approximately
$3.2\,\mbox{KeV}$, whereas its experimentally measured strong shift and width are of order
of $7\,\mbox{eV}$ and $1\,\mbox{eV}$, respectively. Hence,
to a very good approximation, pionic hydrogen can be still considered as a quasi-stable
state. Finally, note that the pion itself decays weakly into the muon and the antineutrino.
The lifetime of this decay is however very large and, therefore, this effect can be safely
neglected.

To briefly summarize, the pionic hydrogen behaves almost as an ordinary hydrogen, but
only almost. Its ground-state energy is shifted and it is not absolutely stable anymore.
The energy shift and width are small as compared to the binding energy and
can be thus treated perturbatively. Measuring these allows one to
determine the pion-nucleon scattering lengths, because both the strong shift and
width are mainly determined by these quantities. In order to achieve this goal, however, one has 
to answer the following questions first:
\begin{enumerate}

\item
  As follows from the discussion above, a natural way to describe pionic hydrogen is
  to assume that, in the first approximation, it represents a stable purely Coulomb
  bound state. The effects of strong interactions should be then included perturbatively.
 What is a consistent framework, based on QCD+QED, which allows one to do so?
 Is it possible to get a systematic expansion of the hadronic atom observables
 in such a theory directly in terms of those observables from the scattering sector
(e.g., scattering lengths), which one aims to eventually extract from the experiment?
  
\item
  In Nature, both strong and electromagnetic interactions contribute to the
  pion-nucleon scattering.
  On the other hand, in order
  to compare with theoretical calculations, one needs scattering lengths and other
  hadronic observables defined in an ideal world with no electromagnetic interactions
  and with equal up- and down-quark masses. How is this ideal world defined, what are
  its parameters and how are the experimental results ``purified'' with respect to
  all kinds of the isospin-breaking effects?

  \end{enumerate}

Here, we shall  give answers to the above questions.

\section{Past and ongoing experiments, physics case}

The experiments on pionic hydrogen and pionic deuterium have been
carried out at PSI for decades~\cite{Hennebach:2014lsa,Hirtl:2021zqf,Strauch:2010vu}, and
have resulted in a very accurate determination of the strong energy shift
and width for these systems. DIRAC experiment 
at CERN has measured the decay lifetime of the ground state and the first excited
state of the $\pi^+\pi^-$ atom (the pionium)~\cite{Adeva:2011tc,DIRAC:2018xvz}, as
well as of the ground state of the $\pi K$ atom~\cite{DIRAC:2017hmz}.
The DEAR and SIDDHARTA experimental collaborations have performed an
accurate measurement of the kaonic hydrogen energy shift and width~\cite{DEAR:2005fdl,SIDDHARTA:2011dsy},
while the SIDDHARTA2 collaboration is measuring, in addition, the energy shift of the kaonic
deuterium~\cite{Curceanu:2023emi}. Furthermore, a wide
range of antiprotonic bound states is studied in different physics laboratories
worldwide~\cite{Doser:2022tlg}. Last but not least, it was recently proposed to study
the properties of the electromagnetic bound states of $DD^*$~\cite{Zhang:2020mpi}
and $DD$ mesons~\cite{Shi:2021hzm}, which may provide very important
clue about strong dynamics in the charm sector.

Each of the above experiments pursues different physics program and provides one missing piece
of information in the big puzzle which goes under the name of the hadron physics at
low energy. The format of the present contribution does not allow us to focus on the physics
case for each experiment in any detail. We shall therefore briefly consider only the
experiments with Goldstone bosons (pions, kaons), as theoretically the cleanest ones.
It is well known that, in the limit of vanishing quark masses, the
self-interactions of the Goldstone bosons, as well as their interactions with other particles,
vanish at zero momenta. This implies vanishing of the pertinent scattering lengths.
Therefore, accurate values of the scattering lengths, which are extracted in the experiments on
hadronic atoms, may render important insight into the mechanism of chiral symmetry breaking
in QCD. Namely, the knowledge of $\pi\pi$ and $\pi K$ scattering lengths, which are determined in the DIRAC
experiment, provides a clue on the size of the quark condensate and helps to decide,
which particular scenario of the {\em spontaneous} chiral symmetry breaking in realized in Nature.
Furthermore, $\pi N$ and $KN$ scattering lengths, which are measured in the Pionic Hydrogen/Pionic Deuterium
and DEAR/SIDDHARTA experiments, respectively, serve as an input in the equations that determine
pion-nucleon and kaon-nucleon $\sigma$-terms and therefore characterize the effect of {\em explicit}
breaking of chiral symmetry due to quark masses in the two- and three-flavor sectors.
In addition, knowing $\sigma$-terms,
it becomes possible to evaluate the so-called strangeness content of the nucleon --
a quantity that is related to the probability of finding virtual $s\bar s$ pairs within the nucleon.

As mentioned before, in order to perform all these wonderful tests of QCD, one should have
a theoretical framework that allows one to accurately extract the scattering lengths from the measured
values of the energy level shift and width. Such a framework has emerged long time ago, within
the potential scattering theory~\cite{Deser:1954vq}. 
The original formula by Deser, Goldberger, Baumann and
Thirring (DGBT) for the ground state of the pionic hydrogen looks particularly simple:
\eq\label{eq:Deser}
\Delta E-\frac{i}{2}\,\Gamma=-\frac{2\pi}{\mu_c}\,|\Psi(0)|^2 \mathscr{A}_c\, .
\en
Here, $\Delta E$ and $\Gamma$ stand for the strong energy shift and the width, respectively (see Fig.~\ref{fig:titlepage}),
the quantity $|\Psi(0)|^2=\alpha^3\mu_c^3/\pi$ denotes the square of the Coulomb wave function
of the $\pi^-p$ pair at the origin, and $\mathscr{A}_c$ is the amplitude
of the $\pi^-p$ elastic scattering at threshold (an exact definition will be given below). This amplitude is complex -- its imaginary part
is determined by the unitarity in a usual way. For example, neglecting the electromagnetic
transitions for a moment, the only sub-threshold channel contains $\pi^0$ and the
neutron. Unitarity then gives:
\eq
\mbox{Im}\,\mathscr{A}_c=2p^*|\mathscr{A}_x|^2\, ,
\en
where $p^*$ denotes the relative momentum of the $\pi^0n$ pair in the CM frame at the $\pi^-p$ threshold, and $\mathscr{A}_x$ is the threshold amplitude
of the process $\pi^-p\to \pi^0n$.
Furthermore, in the isospin limit, both $\mathscr{A}_c$ and $\mathscr{A}_x$ are given
by linear combinations of the S-wave pion-nucleon scattering lengths.
Thus, measuring both $\Delta E$ and $\Gamma$, one may directly extract these scattering lengths.

The formula given above is exact up to the isospin-breaking corrections proportional to $\alpha$
or the quark mass difference $m_d-m_u$. Numerous efforts have been undertaken
since 1954 to improve the accuracy of the original formula and to include these corrections,
which are needed to match the experimental precision. To make the text more readable, we have
moved the references to the most important earlier work to the end. Still, we feel that,
due to the limitations of the present format, it was not possible to fully credit all previous work  
on the subject and refer to the Ref.~\cite{Gasser:2007zt} for a more extensive review.  
Here, we would just like to mention that a qualitative breakthrough in these efforts has been
achieved first by the use of NREFT for the description of hadronic
atoms~\cite{Gall:1999bn,Gasser:2001un}. This method has allowed to set a framework for
a systematic calculation of {\em all} next-to-leading order corrections to the energy shift and
width within ChPT. Next-to-next-to leading order corrections are very small and can be neglected
from the beginning.

In the present work, we shall give a pedagogical introduction to the NREFT,
considering various aspects of this framework in detail. We next discuss the
application of NREFT for the calculation of the spectrum of hadronic atoms.
To this end, we concentrate on
{\em only one} particular example, namely, the ground state of the
pionic hydrogen, and restrict
ourselves to the theoretical background of the problem rather than phenomenological
implications.

Last but not least, we would like to note a striking similarity of the physics of hadronic atoms
to other problems which belong to very different fields of research. All these problems have one in common:
one aims at the understanding of the short-range physics in a system, trapping it in a volume much larger
than the effective radius of the {\em unknown} short-range force, and studying the effect of the latter on the
energy levels. It is crucial that the long-range force, in difference to the short-range interaction,
is {\em exactly known} that allows to uniquely identify (small) effects on the spectrum due to the short-range
interactions and apply the perturbation theory. One example of such systems are the cold atoms
trapped in the harmonic oscillator potential well. The energy level shift, caused by the interaction of the atoms,
is also described by the DGBT formula with the Coulomb wave function replaced by the oscillator wave
function~\cite{Busch:1998cey}. In a close analogy, the trapping of the nucleons by an oscillator potential
on the lattice was proposed in Ref.~\cite{Luu:2010hw}. The aim of this proposal was to study the
$NN$ scattering phase shift by measuring the energy levels. Last but not least, the famous L\"uscher
method to study hadron scattering on the lattice~\cite{Luscher:1986pf,Luscher:1990ux} implies
confining these hadrons in a rectangular box. The role of a trap in this case are played by the boundary
conditions, and the wave function $\Psi(\bm{x})$ describes
the plane waves inside the box. The NREFT method which was used to describe hadronic atoms
works, without much adjustment, in all these cases as well.

\section{How to switch off electromagnetic interactions in QCD+QED?}
\label{sec:splitting}

Let us assume that Nature is described by QCD+QED, whereas other forces are too weak and can be safely
neglected.
The parameters of this theory -- the strong coupling constant $\alpha_S$, the electric charge $e$ and
the quark and lepton masses -- are well-defined at a given renormalization scale $\mu$ (the $\overline{\mbox{MS}}$
renormalization scheme is assumed). One could determine these quantities unambiguously, for example,
by matching to the physical masses and to the fine-structure constant $\alpha$.
Let us now define the ``pure QCD'' world by discarding photon and lepton fields in
the Lagrangian, and setting $m_u=m_d$. 
In the experiments on hadronic atoms, one aims at the extraction of
scattering lengths, which are evaluated in this world. These quantities
are already
``purified'' with respect to all electromagnetic corrections and
isospin-breaking effects due to the up- and down-quark
mass difference.
Note also that, since most calculations in hadron physics
are carried out in this isospin-symmetric world, the experimental results can be meaningfully compared to the results
of theoretical calculations only after such a purification.
It remains to be seen, however, how this goal could be achieved in a consistent manner.

It is important to realize that, unlike the real world, the pure QCD world is not well defined, because its
parameters cannot be unambiguously determined by matching to the experimental input (since the experimental
input from pure QCD does not exist). Thus, pure QCD is a convention.\footnote{By
the way, the same problem arises in lattice calculations (see, for example, the discussion in the recent
FLAG review~\cite{FlavourLatticeAveragingGroupFLAG:2021npn}).} From the viewpoint of the fundamental
theory, the most straightforward way to define the parameters of pure QCD is to declare them equal to the
parameters of QCD+QED at some scale $\mu=\mu_1$~\cite{Gasser:2003hk} (in order to ensure isospin symmetry,
the $u$- and $d$-quark masses are in addition set equal to a common value $\frac{1}{2}\,(m_u+m_d)$ at the
same scale). In this scheme, the ambiguity of defining pure QCD is encoded in the dependence on the matching scale $\mu_1$. Namely, the renormalized masses and couplings in pure QCD and QCD+QED obey different renormalization group equations and hence, choosing different matching
scale $\mu=\mu_2$, the results will change. By the way, it is also seen that
classifying the isospin-breaking corrections into the ``electromagnetic'' and ``strong''
isospin breaking (the latter being attributed to $m_d\neq m_u$) is convention-dependent, since $m_u$ and $m_d$ have different renormalization-group running, when electromagnetic interactions are switched on.

In practice, it is more convenient to use a definition
of pure QCD it terms of physical variables, fixing three fundamental constants of pure QCD, the
strong coupling constant and the light/strange quark masses, in terms of e.g., the proton and charged pion/kaon
masses. In this case, one can straightforwardly use ChPT with virtual photons and
leptons to evaluate isospin-breaking corrections in various amplitudes.
Furthermore, different conventions can be in principle, related to each other
and are equally legitimate. Comparing results of different
calculations, it is important to ensure first that both calculations have used the same
convention.

The above discussion sets stage for a systematic evaluation of the isospin-breaking
corrections within ChPT in the scattering amplitudes extracted from the hadronic
atoms (see Eq.~(\ref{eq:Deser})). ChPT itself, as a relativistic field theory, is however not
well suited to deal with the bound states like hadronic atoms, and new ideas are
necessary to make this very complicated problem tractable.

\section{Hadronic atoms in the NREFT framework}

\subsection{Essentials}

\begin{figure}[t]
\begin{center}
  \includegraphics[width=14.cm]{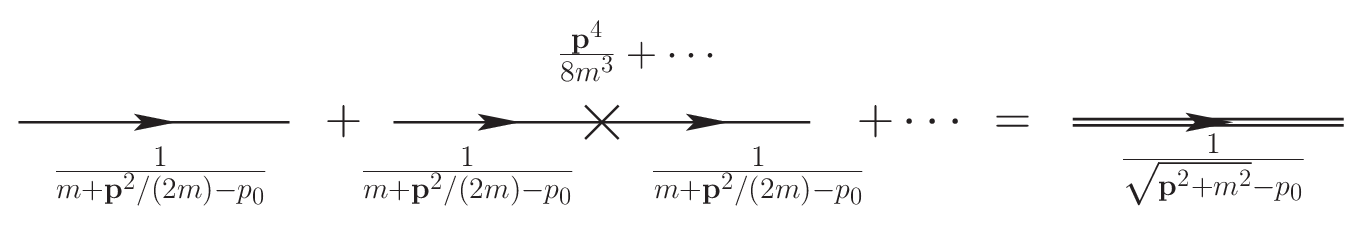}
\caption{Relativistic insertions in a free particle propagator.}
  \label{fig:propagator}
\end{center}
  \end{figure}

In their seminal paper, Caswell and Lepage~\cite{Caswell:1985ui} have first introduced
NREFT for the description of QED bound states. Since then, the approach has proven
extremely useful in many fields of physics. Among these applications, we mention only
NRQCD~\cite{Brambilla:2004jw}, the study of unitary cusp in the three-particle
decays~\cite{Gasser:2011ju} and the three-particle quantization condition on the
lattice~\cite{Hammer:2017kms}. The foundations of the method are very transparent.
A fundamental assumption that lies at the foundation of the non-relativistic effective theory
is that the typical three-momenta of particles in the system are much smaller than the
masses of these particles. This assumption works well in case of hadronic atoms, where
typical momentum $p\sim \alpha\mu_c$ is given by the inverse of the Bohr radius.
Moreover, parametrically, the typical momentum is proportional to the fine-structure constant
$\alpha$ and the non-relativistic expansion translates into a systematic expansion of the observables of a system into the powers of $\alpha$, setting
power-counting rules in the NREFT. Furthermore, in order to create a
particle-antiparticle pair in a relativistic theory, one needs momenta of order $m$,
where $m$ denotes the mass of a particle. If the typical momenta
are of order of $\alpha\mu_c$, these two scales are well separated and the effects of the antiparticles can be systematically included in the couplings of the effective Lagrangian, leading to a theory where annihilation/pair creation is forbidden. Therefore, the particle number is conserved. In practical terms, this can be realized in the following manner.
Consider, for definiteness, a theory with one real scalar field with a mass $m$.
The free propagator in this theory is given by
\eq
D(p^2)=\frac{1}{m^2-p^2-i\varepsilon}=\frac{1}{2w(\bm{p})}\, 
\biggl\{\frac{1}{w(\bm{p})-p_0-i\varepsilon}+\frac{1}{w(\bm{p})+p_0-i\varepsilon}
\biggr\}\, ,\quad\quad
w(\bm{p})=\sqrt{m^2+\bm{p}^2}\, .
\en
The second term in the above equation, which corresponds to the contribution of an
antiparticle, can be expanded in powers of $(p_0-w(\bm{p}))$ and $\bm{p^2}/m^2$.
The whole contribution, coming from this term, can be eventually included in the couplings
of the effective Lagrangian. In the non-relativistic theory, the propagator is given by the
first term of the above equation.  The Lagrangian, which produces this propagator, is linear in the time derivative.
In order to ease the notation, it is convenient (but not mandatory)
to further rescale the field
$\phi(x)\to (2w(-i\nabla))^{-1/2}\phi(x)$. Then, the free non-relativistic Lagrangian takes the form
\eq
\mathscr{L}_{\sf free}(x)=\phi^\dagger(x)\left(i\partial_t-w(-i\nabla)\right)\phi(x)
=\phi^\dagger(x)\left(i\partial_t-m+\frac{\nabla^2}{2m}+\frac{\nabla^4}{8m^3}+
  \cdots\right)\phi(x)\, .
\en
At this place, one has a freedom of the choice of the non-interacting Lagrangian. One could either regard the whole expression as an unperturbed Lagrangian, or shift the terms
containing $\nabla^4$ and higher into the interaction part. Each of the choices has is advantages and drawbacks. Below, we stick to the second path, in which the unperturbed Lagrangian takes the form\footnote{
In general, the non-relativistic theories differ from the relativistic ones in two aspects.
The {\em dynamical} aspect that implies integrating out antiparticles and the
{\em kinematical} aspect boils down to replacing the relativistic dispersion relation
$p_0=w(\bm{p})$ by its non-relativistic counterpart $p_0=E(\bm{p})$ from Eq.~(\ref{eq:phi}). The former
is essential and in fact defines a non-relativistic theory. The latter is optional
and can be circumvented -- for example, one may write down the NREFT in a fully
covariant form, see~\cite{Gasser:2011ju}. The problems with the power counting,
which emerge in this approach, can be solved by the use of the so-called threshold
expansion of the Feynman integrals.}

\eq
\mathscr{L}_0(x)=\phi^\dagger(x)\left(i\partial_i-m+\frac{\nabla^2}{2m}\right)\phi(x)\, .
\en
Here, $\phi(x)$ is a non-relativistic field that contains only annihilation operator
(we remind the reader that the equation of motion for this field contains only the first-order
derivative in time):
\eq\label{eq:phi}
\phi(x)=\int\frac{d^3\bm{p}}{(2\pi)^3}\,e^{-ipx}a(\bm{p})\, ,\quad\quad
p^\mu=(E(\bm{p}),\bm{p})\,,\quad E(\bm{p})=m+\frac{\bm{p}^2}{2m}\, .
\en
The creation/annihilation operators obey commutation relations
\eq
[a(\bm{p}),a^\dagger(\bm{k})]=(2\pi)^3\delta^3(\bm{p}-\bm{k})\, ,
\en
and the propagator is given by
\eq
i\langle 0|T\phi(x)\phi^\dagger(y)|0\rangle=\int\frac{d^4p}{(2\pi)^4}\,\frac{1}{E(\bm{p})-p_0-i\varepsilon}\, .
\en
The higher-order insertions into the two-point function, produced by the 
operators $\phi^\dagger \dfrac{\nabla^4}{8m^3}\,\phi$, \ldots can be summed up to all orders and ensure that the pole of the propagator is at the right
position $p_0=w(\bm{p})$ (see Fig.~\ref{fig:propagator}).

\subsection{Non-relativistic Lagrangian}

Below, we shall first concentrate on the hadronic sector of the theory and consider the
inclusion of the electromagnetic interactions at a later stage. Furthermore,
in order to avoid a clutter of
indices, at the first step we restrict ourselves to the case of a single massive scalar field. 
The non-relativistic Lagrangian is given by an infinite tower of terms
with increasing mass dimension. The guiding principles for the construction of the
Lagrangian are listed below:

\begin{itemize}
\item
  The interaction vertices in the Lagrangian conserve particle number. For example,
  the vertices of the type $\phi^\dagger\phi(\phi+\phi^\dagger)$ are not allowed.
  Owing to the particle number conservation, the Hilbert space falls apart into sectors
  containing different number of massive particles. These sectors do not talk to each other.
  For example, the vertex of the type $\phi^\dagger\phi^\dagger\phi^\dagger\phi\phi\phi$
  might be present but remains invisible from the two-particle sector. Since we are studying
  hadronic atoms that consist of two particles, such vertices can be merely
  ignored. Note, however that, to the opposite,
  the dynamics in the three-particle sector does depend on
  the interactions in its two-particle subsystems.

\item
  The building blocks of the Lagrangian should be constructed, according to certain
  symmetry principles. Namely, the Lagrangian should be rotationally invariant
  (explicit Lorentz-invariance is lost in the formulation we are using and re-emerges only
  at the level of the effective couplings, which obey certain relations that reflect
  Lorentz-invariance of the underlying theory). The underlying
  space-time discrete symmetries $P,T$ should also persist in the non-relativistic theory.
  The case with charge conjugation $C$ is more subtle, because antiparticles are integrated
  out. These could be restored as independent degrees of freedom (for example, the electron and positron are described by separate fields in NRQED). The $C$-invariance then requires
  that the effective couplings in the particle and antiparticle sectors are the same.
  When electromagnetic interactions are turned on, the Lagrangian, in addition, should be gauge-invariant.

\item
  In general, the NREFT Lagrangian in not Hermitian. The imaginary part of the coupling constants encode subthreshold contributions from the channels that are integrated out.
  For example, the threshold for the process $\pi^-p\to n\gamma$ lies below
  of $\pi^-p$ threshold. One could integrate out the former because the mass
  gap $\sim M_\pi$ is much larger that the typical momentum $\alpha\mu_c$. The imaginary
  part of the effective coupling that describes the transition  $\pi^-p\to \pi^-p$
  becomes non-vanishing, in order to satisfy the unitarity. This case
  is discussed below in detail. Note also that the time invariance implies
  $T\mathscr{L}(x)T^{-1}=\mathscr{L}^\dagger(-x_0,\bm{x})$. Hence, the $T$-invariance
  may still hold in a theory described by a Lagrangian with complex couplings.

\item
  Counting rules lie at heart of any EFT. In our case, a generic small
  parameter is given by $|\bm{p}|/m\sim\alpha=O(\delta)$. The different building
  blocks in the Lagrangian contain different number of space derivatives and can be
  therefore ordered according to the above power counting (the time derivatives
  can always be eliminated with the use of the equations of motion). These counting rules
  hold at tree level but, in general, could be upset by loop corrections. Applying the
  so-called threshold expansion, it is possible to restore power counting. We shall consider
  this procedure below in more detail.

  \end{itemize}

  Equipped with this knowledge, we may proceed with writing down few terms in the
  NREFT Lagrangian. In the purely strong sector, the Lagrangian containing
  a single scalar field takes the following form:
  \eq\label{eq:L}
  \mathscr{L}=\phi^\dagger\left(i\partial_t-m+\frac{\nabla^2}{2m}\right)\phi
  +\phi^\dagger\left(\frac{\nabla^4}{8m^3}+\cdots\right)\phi(x)
  +c_0\phi^\dagger\phi^\dagger\phi\phi
+c_2\left((\phi^\dagger\lrarrow^2\phi^\dagger)(\phi\phi)+\mbox{h.c.}\right)+\cdots\, ,
  \en
  where $a\lrarrow b\doteq a\nabla b-b\nabla a$
  and $a\lrarrow^2b=a\nabla^2b+b\nabla^2a-2\nabla a\nabla b$.
  The dots denote terms containing higher
  number of space derivatives (the time derivatives can always be removed by using
  equations of motion). In order to maintain the proper dimension, the couplings $c_i$ 
  should contain powers of the heavy mass scale in the denominator (in the non-relativistic theories, the role of the heavy scale is played by the mass $m$ itself). Note also that here one assumes the couplings to have natural size, i.e., no shallow bound states in the purely strong sector are present.

  \subsection{Loops}

\begin{figure}[t]
  \begin{center}
    \includegraphics*[width=14.cm]{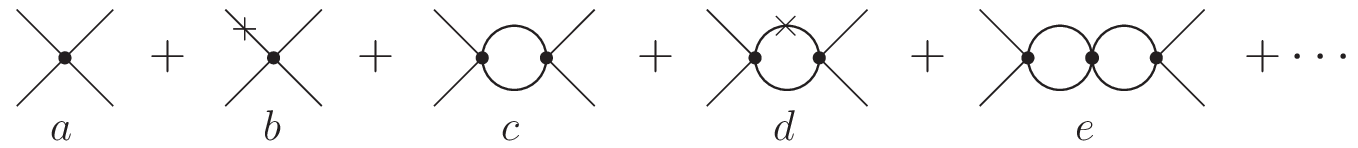}
  \caption{Bubble graphs contributing to the non-relativistic two-particle scattering
    amplitude. The filled circle denotes the local four-particle vertex, 
    and crosses represent the relativistic insertions in the external and internal
    lines.}
  \label{fig:bubblechain}
  \end{center}
  \end{figure}

  It is crucial to realize that the structure of the Feynman diagrams in the non-relativistic
  theories simplifies dramatically as compared to the relativistic ones.
    Out of all possible diagrams, only
  the bubble chain shown if Fig.~\ref{fig:bubblechain} survives while, as a direct consequence of the particle number conservation, all $t$-, $u$-channel bubbles, as well as the diagrams with multiparticle intermediate states, etc., vanish. Furthermore, let us consider the single loop
  shown on this figure. We work in the CM frame and denote the total
  momentum if the incoming/outgoing pair by $P^\mu=(P_0,\bm{0})$. We further
  assume, for simplicity,  that only the lowest-order coupling $c_0\neq 0$.
The expression
  for the loop with no derivative vertices is given by
  \eq
  J(q_0)=\int \frac{d^Dk}{(2\pi)^Di}\,\frac{1}{(E(\bm{k})-k_0-i\varepsilon)
    (E(\bm{k})-P_0+k_0-i\varepsilon)}
  =\int\frac{d^dk}{(2\pi)^d}\,\frac{1}{(2E(\bm{k})-P_0-i\varepsilon)}\, .
  \en
  Here, dimensional regularization is used for the ultraviolet-divergent loop, with $D=d+1$ being the number of space-time dimensions. Evaluating this integral is straightforward,
  and the result is given by
  \eq\label{eq:J}
  J(q_0)=-\frac{m}{4\pi}\,\sqrt{-m(P_0-2m)-i\varepsilon}=
  -\frac{m}{4\pi}\,\sqrt{-q_0^2-i\varepsilon}\, ,\quad\quad P_0=2m+\frac{q_0^2}{m}\, .
  \en
  Above threshold, where $q_0^2>0$, we have $J(q_0)=imq_0/(4\pi)$.
  Note that the result is finite in dimensional regularization and does not require further
  subtractions. Hence, the non-relativistic on-shell scattering amplitude to all orders is given in a form of the geometric series
  \eq\label{eq:TNR0}
  T_{\sf NR}(q_0)=(4c_0)+(4c_0)^2\frac{1}{2}\,J(q_0)+(4c_0)^3\left(\frac{1}{2}\,J(q_0)\right)^2+\cdots=\frac{1}{(4c_0)^{-1}-\dfrac{1}{2}\,J(q_0)}\, .
  \en
  This result can be easily generalized to include derivative couplings.
  Let us limit ourselves
  first to the S-wave scattering only.
  At tree level, the off-shell scattering amplitude can be directly read
  off from the Lagrangian
  \eq
  \langle \bm{p}|T_{\sf NR}^{(0)}(q_0)|\bm{q}\rangle=4c_0-16c_2(\bm{p}^2+\bm{q}^2)\, .
  \en
 Furthermore, the one loop diagram is given by the expression
 \eq\label{eq:TNR1}
  \langle \bm{p}|T_{\sf NR}^{(1)}(q_0)|\bm{q}\rangle
  =\frac{1}{2}\,\int\frac{d^dk}{(2\pi)^d}\,\frac{(4c_0-16c_2(\bm{p}^2+\bm{k}^2))(4c_0-16c_2(\bm{k}^2+\bm{q}^2))}{(2E(\bm{k})-P_0-i\varepsilon)}\, ,
  \en
  and so on. Furthermore,
  in the numerator, one can replace $\bm{k}^2$ by  $q_0^2$, since in the difference the
  denominator cancels, leading to the no-scale integral which is known to vanish in dimensional regularization. In general, within dimensional regularization, one can consistently pull out
  the factors $(\bm{k}^2)^n$ out of all integrals. Going to the mass shell
  $\bm{p}^2=\bm{q}^2=P_0^2/4-m^2=q_0^2+O(q_0^4)$, we finally get the following expression to all orders
  \eq
  T_{\sf NR}(q_0)=\frac{1}{R^{-1}(q_0)-\dfrac{1}{2}\,J(q_0)}\, ,\quad\quad
  R(q_0)=4c_0-32c_2q_0^2+\cdots\, .
  \en
  A similar approach can be used for the higher partial waves. Finally, above threshold,
  the self-energy insertion in a single loop yields the following result:
  \eq\label{eq:J1}
  J_1(q_0)=2\int \frac{d^Dk}{(2\pi)^Di}\,\frac{1}{(E(\bm{k})-k_0-i\varepsilon)^2
    (E(\bm{k})-P_0+k_0-i\varepsilon)}\,\frac{\bm{k}^4}{8m^3}
  =\int\frac{d^dk}{(2\pi)^d}\,\frac{1}{(2E(\bm{k})-P_0-i\varepsilon)^2}
  \,\frac{\bm{k}^4}{4m^3}=\frac{iq_0}{4\pi}\left(-\frac{5q_0^2}{8m^2}\right)\, ,
  \en 
  and so on.

  The above examples exhaust all possible types of diagrams in the strong sector.
  To summarize, the most important conclusion is that the power counting is not
  broken by the loops in the strong sector
  when the dimensional regularization is used. As seen, the
  higher-order insertions result in the factors
  $(q_0^2/m^2)^n$ in front of the result.
  
  This observation can be put on a more formal footing, using the power counting
  rules introduced above.
  Namely, one counts all three-momenta as $O(\delta)$, where $\delta$ denotes a generic
  small parameter ($q_0=O(\delta)$, of course). The kinetic energy
  $p_0-m$ is counted as $O(\delta^2)$. The lowest-order non-derivative vertex is $O(\delta^0)$
  and each space derivative adds $O(\delta)$ to it. Furthermore, according to our counting,
  each propagator is $O(\delta^{-2})$ and the four-dimensional integration results in
$dp_0d^3\bm{p}\sim O(\delta^2\cdot \delta^3)=O(\delta^5)$. Owing to this counting, 
  the leading tree-level contribution in Eq.~(\ref{eq:TNR0}), containing $c_0$, counts
  as $O(\delta^0)$, the leading one-loop contribution in Eq.~(\ref{eq:TNR1}) counts
  as $O(\delta)$, and so on. Contributions with $n$ space derivatives in the vertices in total
  are down by $\delta^n$ with respect to the leading contribution. Finally, an insertion of
  $\bm{k}^4$ in the propagator produces one additional propagator. Hence this
  diagram counts as $O(\delta^4\cdot \delta^{-2})=O(\delta^2)$ with respect to the diagram with no
  insertion. This is exactly the scaling in $q_0$ that we observe in all examples considered
  above. To summarize, the naive counting rules, established at the level of the
  effective Lagrangian, are valid to all orders in perturbation theory, if dimensional
  regularization plus minimal subtraction is used in the loop. If another type of regularization
  (e.g., cutoff regularization) is used, the naive counting rules will be, in general, violated.

  \subsection{Matching}

  The non-relativistic Lagrangian~(\ref{eq:L}) contains unknown couplings
  $c_0,c_2,\ldots$ If this Lagrangian were to describe QCD at low energy, these
  couplings had to be expressed, in principle, in terms of the parameters of QCD (in fact, the
  same statement holds for an arbitrary underlying field theory). The procedure
  described above goes under the name of matching and states that the $S$-matrix
  elements, calculated in the underlying and effective theories, are the same up to a given
  order. Specifically, in our case, the matching condition is formulated in terms of
  the two-body scattering amplitude. Recalling that the normalization of the one-particle
  states in the relativistic and non-relativistic theories differs by a kinematic factor
  $2w(\bm{p})$, the matching condition for the scattering amplitudes of the process
  $q_1+q_2\to p_1+p_2$ takes the form
  \eq\label{eq:matching}
  \prod_{i=1}^2(2w(\bm{p}_i))^{1/2}(2w(\bm{q}_i))^{1/2} T^{\sf NR}(\bm{p}_1,\bm{p}_2;\bm{q}_1,\bm{q}_2)= T^{\sf R}(\bm{p}_1,\bm{p}_2;\bm{q}_1,\bm{q}_2)\, .
  \en
  Here, $p_{i0}=\sqrt{m^2+\bm{p}_i^2}$ and  $q_{i0}=\sqrt{m^2+\bm{q}_i^2}$
for $i=1,2$.
  
  Next, according to Eq.~(\ref{eq:TNR0}), the expansion of the non-relativistic amplitude
  near the elastic threshold takes the form
  \eq
  T^{\sf NR}(\bm{p}_1,\bm{p}_2;\bm{q}_1,\bm{q}_2)=4c_0+\frac{2imc_0^2}{\pi}\,q_0+
  O(q_0^2)\, ,\quad\quad q_0^2>0\, .
  \en
  The important property of this relation is that it is valid to all orders in perturbation theory.
  Indeed, as seen, the inclusion of the derivative vertices, as well as kinematic insertions
  into the internal propagators lead to the contributions containing higher powers of $q_0$.
  Hence, the first two coefficients in this expansion are determined by a single coupling
  constant $c_0$ and will not be modified by perturbative corrections. The
  coupling $c_2$ arises first at $O(q_0^2)$. Recall
  now that the relativistic amplitude at threshold, by definition, is given by the S-wave
  scattering
  length $a$ modulo kinematic factor. Then, from matching, a simple relation follows, valid to all orders {\em both} in the NREFT and underlying theory:
  \eq
  c_0=-\frac{2\pi a}{m}\, .
  \en
As seen from the above relation,
the parameters of the NREFT are expressed in terms of the {\em observables}
in the underlying theory (for example, the scattering length). Hence, analyzing experimental data in the NREFT
framework, one can directly extract these
observables without resorting to the perturbative
expansion in the underlying relativistic theory (ChPT, in our case). It should be 
realized that this is a crucial property of the NREFT that allows to determine
the scattering lengths
from the experiments on hadronic atoms very accurately.
Putting differently, NREFT allows to effectively subsum all
relevant contributions of ChPT near threshold and parameterize them in terms of
observables. Further, it allows to directly establish a relation between hadronic atom 
spectrum and these observables.

The picture remains practically the same at higher orders in the expansion in $q_0$. In fact,
non-trivial matching conditions arise at even orders in $q_0$, since odd orders are fixed
by unitarity. At $O(q_0^2)$,  the new coupling $c_2$ enters, which is determined by the
S-wave scattering length $a$ and the effective range $r$. At this order, a relativistic
correction that emerges from the kinematic insertion into the internal lines, also
contributes. At higher orders, tracking these relativistic corrections order by order
becomes too complicated, whereas a covariant formulation along the lines of
Ref.~\cite{Gasser:2011ju} provides a relatively simple framework. However, as going
to the higher orders will not be needed here, we do not further elaborate on this issue.

\subsection{Including photons}

In order to describe hadronic atoms, one has to construct NREFT including both hadrons and
photons. Constructing the Lagrangian, based on the gauge invariance,
is a straightforward task. For demonstration,
consider a pair of oppositely charged non-relativistic scalar particles $\phi_\pm(x)$
and write down
the Lagrangian using gauge-covariant building blocks: the covariant
derivative $D_t\phi_\pm=\partial_t\phi_\pm\mp ieA_0\phi_\pm$,
$\bm{D}\phi_\pm=\nabla\phi_\pm\pm ie\bm{A}\phi_\pm$, as well as electric
$\bm{E}$ and magnetic $\bm{B}$ fields:
\eq\label{eq:L-em}
\mathscr{L}=\sum_\pm\phi_\pm^\dagger\left(iD_t-m+\frac{\bm{D}^2}{2m}
+\frac{\bm{D}^4}{8m^3}+\cdots
\mp \frac{eh_1}{6m^2}\,(\bm{D}\bm{E}-\bm{E}\bm{D})+\cdots
\right)\phi_\pm+c_0\phi_+^\dagger\phi_-^\dagger\phi_+\phi_-+\cdots-\frac{1}{4}\,F_{\mu\nu}F^{\mu\nu}\, .
\en
Here, $F_{\mu\nu}=\partial_\mu A_\nu-\partial_\nu A_\mu$ is the electromagnetic
field strength tensor. In order to ease the notations, we have included only one quartic coupling,
describing the elastic scattering of the oppositely charged particles. All other
reaction channels can be considered in the similar fashion. Note also that, as compared
to the purely strong case, a new effective coupling, $h_1$, appears in the one-particle
sector. It can be easily seen that this coupling can be expressed through the
mean square radius of a charged particle:
\eq
h_1=m^2\langle r^2\rangle\, .
\en
Owing to the $C$-invariance, there is a single coupling $h_1$ for positively and negatively
charged particles. By the same token,
the quartic couplings in the channels that are obtained
by charge conjugation (not shown explicitly in Eq.~(\ref{eq:L-em})), are also the same.
Of course, inclusion of all charge-conjugated channels in a non-relativistic theory is
optional -- if some of these are dropped, the theory is still consistent.

In order to quantize the theory described by the Lagrangian~(\ref{eq:L-em}), one should
fix the gauge. Here, it is worth mentioning that the choice of gauge is not linked to the
one in the underlying theory. Indeed, two theories are connected by the matching
condition that always involves gauge-independent {\em observables.} One may use
this freedom and work in a non-covariant Coulomb gauge, which turns out to be the most
convenient one in the non-relativistic calculations:
\eq
D^{\mu\nu}(k)&=&i\int d^4x e^{ikx}\langle 0|A^\mu(x)A^\nu(0)|0\rangle\, ,
\nonumber\\[2mm]
D^{00}(k)&=&-\frac{1}{\bm{k}^2}\, ,\quad\quad
D^{ij}(k)=-\frac{1}{k^2+i\varepsilon}\,\left(\delta^{ij}-\frac{k^ik^j}{\bm{k}^2}\right)\, ,
\quad\quad
D^{i0}(k)=D^{0i}(k)=0\, .
\en
In the following, we shall refer to the photons carrying index $0$ and $i$, to as the Coulomb and transverse photons, respectively. Note that the Coulomb photons do not propagate in time -- the propagator containing Coulomb photons is proportional to $\delta(t)$ in position space.

In the perturbation theory involving photons, two new problems arise:

\begin{itemize}
\item The loops involving photons, in general, break the naive power counting established
  above.
\item The matching condition, formulated at threshold in the purely strong case, cannot
  be applied anymore since, owing to the infrared divergences, the threshold scattering
  amplitudes are in general infinite.
  \end{itemize}

  We address the first problem only very briefly and refer, e.g., to the
  textbook~\cite{Meissner:2022cbi} or to the review article~\cite{Gasser:2007zt}
  for more details. The integrands in the Feynman
integrals with transverse photons are not, in general,
homogeneous functions of the parameter $\delta$. 
The standard way to circumvent the resulting violation of the power counting
is provided by the so-called
threshold expansion \cite{Beneke:1997zp}, or expansion by regions \cite{Mohr:2005pv}.
In the context of the present problem, the prescription consists merely in the
identification of all regions with different scaling of the integration momentum $\bm{k}$.
After that, the integrand should be expanded in Taylor series, retaining only the leading-order term in the denominator and, subsequently, the integration should be carried out
in dimensional regularization.
It can be shown that, adding all pieces together, one reproduces the exact result
for the initial integral.
Reenacting power counting by dropping the contribution from the hard regime,
$\bm{k}=O(\delta^0)$, which is a polynomial in the external momenta, amounts to merely
changing  the renormalization prescription and does not alter the physics content.

\begin{figure}[t]
  \begin{center}
    \includegraphics[width=7.cm]{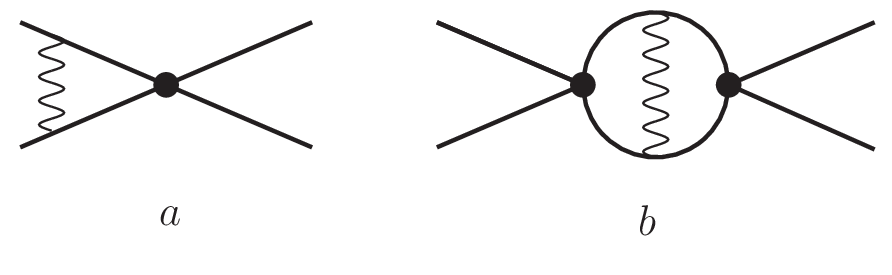}
    \caption{$O(\alpha)$ corrections to the strong amplitude: a) Coulomb photon exchange in the external lines, and b) Coulomb photon exchange in the intermediate state.
      The shaded blob denotes the full 4-particle vertex. All other $O(\alpha)$ corrections
    do not contribute at the accuracy we are working.}
   \label{fig:Coulomb-Oalpha}
  \end{center}
 \end{figure}

The second problem will be addressed in more detail.
In the presence of photons, the threshold amplitude is
infrared-divergent and cannot be used to write down the matching condition without
further ado. Here we consider the matching condition at $O(\alpha)$
from the beginning and neglect all corrections beyond it.
Furthermore, it is straightforward to define a set of one-photon-reducible diagrams, which
contains only diagrams that can be made disconnected by cutting a single photon
line. It is convenient to formulate the matching condition it terms of the one-photon
irreducible amplitude, dropping all reducible contributions.
In Fig.~\ref{fig:Coulomb-Oalpha}, a full set of the infrared-singular irreducible diagrams
at this order, containing the exchange of Coulomb photons, is shown. These include:
(two) diagrams describing photon exchange between initial- or final-state particles
(Fig.~\ref{fig:Coulomb-Oalpha}a), and the photon exchange between particles in the intermediate state (Fig.~\ref{fig:Coulomb-Oalpha}b). An infinite number of purely strong bubbles can be attached to these diagrams. Moreover, since the vertex of the transverse photons is suppressed at $O(\delta)$, the diagrams with the exchange of transverse photons will be suppressed at $O(\delta^2)$ with respect to the corresponding diagrams with Coulomb photon exchanges. Examining carefully all possible topologies of diagrams with transverse
photons, one may ensure that none of them contributes to the matching condition
for the amplitude at the accuracy we are working.

We carry out calculations in the CM frame, where $\bm{p}_1=-\bm{p}_2=\bm{p}$,
$\bm{q}_1=-\bm{q}_2=\bm{q}$ and $p_{10}+p_{20}=q_{10}+q_{20}=P_0$,
$|\bm{p}|=|\bm{q}|=q_0$.
The scattering amplitude, obtained from the diagram 
Fig.~\ref{fig:Coulomb-Oalpha}a, is written down as
\eq\label{eq:Vc}
-c_0V_c(q_0)
&=&e^2c_0\int\frac{d^d\bm{k}}{(2\pi)^d}\,\frac{1}{|\bm{p}-\bm{k}|^2}\,
\frac{1}{2E(\bm{k})-P_0-i\varepsilon}
\nonumber\\[2mm]
&=&c_0\,\frac{\alpha m\mu^{d-3}}{2(-q_0^2-i\varepsilon)}
\left(\frac{1}{d-3}-\frac{1}{2}\,(\Gamma'(1)+\ln 4\pi)+\frac{1}{2}\,\ln\frac{-4(q_0^2+i\varepsilon)}{\mu^2}\right)
+O(d-3)\, , 
\en
where $\mu$ stands for the scale of dimensional regularization.

The two-loop diagram Fig.~\ref{fig:Coulomb-Oalpha}b corresponds to
\eq\label{eq:Bc}
c_0^2B_c(q_0)
&=&e^2c_0^2\int \frac{d^d\bm{l}}{(2\pi)^d}\,\frac{d^d\bm{k}}{(2\pi)^d}\,\frac{1}{2E(\bm{l})-P_0-i\varepsilon}\,\frac{1}{|\bm{l}-\bm{k}|^2}\,\frac{1}{2E(\bm{k})-P_0-i\varepsilon}
\nonumber\\[2mm]
&=&-\frac{\alpha m^2\mu^{d-3}}{8\pi}\,c_0^2
\left(\frac{1}{d-3}-\Gamma'(1)-\ln 4\pi-1+\ln\frac{m^2}{\mu^2}+
\ln\frac{-4(q_0^2+i\varepsilon)}{m^2}\right)
\doteq c_0^2\left(\bar B_c-\frac{\alpha m^2}{8\pi}\,\ln\frac{-4(q_0^2+i\varepsilon)}{m^2}\right)\, .\quad\quad
\en
This exhausts all possible infrared-singular contributions to the irreducible scattering amplitude
$\langle\bm{p}|T^{\sf NR}|\bm{q}\rangle$ at threshold at order $\alpha$.
Putting now all pieces together, one can write
\eq\label{eq:threshold}
(1+2V_c(q_0))\langle\bm{p}|T^{\sf NR}|\bm{q}\rangle
=\tilde{\mathscr{A}}+\tilde{\mathscr{B}}\ln\frac{-4(q_0^2+i\varepsilon)}{m^2}+
O(q_0,\alpha^2)\, ,
\en
where
\eq\label{eq:threshold1}
\tilde{\mathscr{A}}=c_0(1+c_0\bar B_c)\, ,\quad\quad
\tilde{\mathscr{B}}=-\frac{\alpha m^2}{8\pi}\,c_0^2\, .
\en
Thus, both singular and regular pieces of the irreducible amplitude are given in terms of a single constant $c_0$. Note also that the dependence on the
on the scale $\mu$ has disappeared.

The matching condition for the full amplitude is again written as in Eq.~(\ref{eq:matching}), and can be used separately for reducible and irreducible parts. The key property is
that, near threshold, the irreducible amplitude in the underlying theory
must also have the form given by Eq.~(\ref{eq:threshold}).
The threshold amplitude $\mathscr{A}$ in the underlying theory will be equal to
 $\tilde{\mathscr{A}}$ modulo overall normalization.
Hence, as in purely
strong case, the matching condition relates $c_0$ to the threshold amplitude
$\mathscr{A}$ in the underlying theory to all orders in NREFT expansion.

Finally, note that $c_0$ defined in the above manner depends, in general, on
the fine-structure
constant $\alpha$:
\eq
c_0=\bar c_0+\alpha c_0'+\cdots\, ,
\en
where $\bar c_0$ is a constant defined in ``purely strong'' (with $\alpha=0$)
underlying theory, which is expressed through the scattering length
one wants to extract from the experiment (see Sect.~\ref{sec:splitting}).

\subsection{A short intermediate summary}

In this section we considered in detail the essentials of the NREFT matched to an underlying relativistic theory with photons.
The main idea beyond this approach is that the matching in the NREFT Lagrangian
can be carried out in perturbation theory in the fine-structure constant $\alpha$.
The matching condition at lowest order in $\alpha$ is written in terms of
observables (purely strong scattering lengths).

After determining all parameters in the non-relativistic Lagrangian, one can use it for
studying shallow bound states (hadronic atoms) which is an inherently non-perturbative
problem. A crucial simplification comes from the fact that NREFT is in fact equivalent
to the quantum-mechanical Schr\"odinger equation for the bound states. This allows
one to solve the bound-state problem with a surprising elegance and ease.
Below, we shall
describe the calculation of the spectrum in a particular example of pionic hydrogen.
However, the method can be applied, without any substantial modification, to almost all
known hadronic atoms.

\section{Spectrum and decays of the pionic hydrogen}

\subsection{The Lagrangian and matching}

Neglecting weak interactions and assuming nucleons and charged pions to be stable
particles, the pionic hydrogen can decay either into $n\pi^0$ or into the neutron and the
photons. The latter cannot be included explicitly in the NREFT framework, because
the mass gap $\sim M_\pi$ represents a hard scale of the theory (in the following,
$M_\pi$ and $M_{\pi^0}$ denote the charged and the neutral pion masses, whereas
$m_p,m_n$ stand for the proton and neutron masses, respectively).
In case of the $n\pi^0$
channel, one has a choice: either one includes $n$ and $\pi^0$ as explicit degrees of
freedom and considers the two-channel problem, or one integrates out them from the
theory. The price to pay for this is that the role of the heavy scale in this case 
is played by the mass gap $\Delta=m_p+M_\pi-m_n-M_{\pi^0}\simeq 3.3\,\mbox{MeV}$,
much smaller than $M_\pi\simeq 140\,\mbox{MeV}$. Defining now a generic small isospin-breaking parameter $\delta\sim\alpha\sim (m_d-m_u)$, we have
$\Delta=O(\delta)$ and $M_\pi=O(\delta^0)$. Due to this, some effective
couplings will be enhanced (contain negative powers of $\delta$), as compared
to the two-channel picture. We shall prefer working in the one-channel case and show,
how the novel counting rules apply.

The non-relativistic Lagrangian is given by
\eq
\mathscr{L}&=&-\frac{1}{4}\,F_{\mu\nu}F^{\mu\nu}+\psi^\dagger\left(iD_t-m_p
+\frac{\bm{D}^2}{2m_p}+\frac{\bm{D}^4}{8m_p^3}+\cdots
-\frac{ec_p^F}{2m_p}\,\boldsymbol{\sigma}\bm{B}
-\frac{ec_p^D}{8m_p^2}\,(\bm{D}\bm{E}-\bm{E}\bm{D})
-\frac{iec_p^S}{8m_p^2}\,\boldsymbol{\sigma}\cdot(\bm{D}\times\bm{E}
-\bm{E}\times\bm{D})+\cdots\right)\psi
\nonumber\\[2mm]
&+&\pi_-^\dagger\left(iD_t-M_\pi+\frac{\bm{D}^2}{2M_\pi}
+\frac{\bm{D}^4}{8M_\pi^3}+\cdots- \frac{eh_1}{6M_\pi^2}\,
(\bm{D}\bm{E}-\bm{E}\bm{D})+\cdots\right)\pi_-
+c_0(\psi^\dagger\pi_-^\dagger)(\pi_-\psi)
+c_2\left((\psi^\dagger\lrarrowD^2\pi_-^\dagger)(\pi_-\psi)+\mbox{h.c.}\right)+\cdots\, .
\en
Here, $\psi$ and $\pi_-$ stand for the non-relativistic proton and negative pion fields,
respectively, and $D_t\psi=\partial_t\psi-ieA_0\psi$,
$\bm{D}\psi=\nabla\psi+ie\bm{A}\psi$.
The various coupling constants entering the above expression are determined by using matching condition for the pion and proton electromagnetic form factors,
as well as the $\pi^-p$ scattering amplitude. In the one-particle sector, one obtains
\eq
c_p^F=1+\kappa_p\, ,\quad\quad
c_p^D=1+2\kappa_p+\frac{4}{3}\,m_p^2\langle r_p^2\rangle\, ,\quad\quad
c_p^S=1+2\kappa_p\, ,\quad\quad
h_1=M_\pi^2\langle r_\pi^2\rangle\, .
\en
Here, $\langle r_p^2\rangle,\,\langle r_\pi^2\rangle$ denote the charge radius squared
obtained from the proton and the pion form factors, and $\kappa_p$ is the anomalous
magnetic moment of a proton.

The matching in the two-particle sector is more subtle. To start with, according to
our power counting, the relative three-momentum of the charged pair in the bound state
should be of order $\delta$ and, therefore, the effective-range term with $c_2$ should
not contribute. This conclusion is however premature, since a small energy
scale $\Delta=O(\delta)$ was integrated out. As a result of this, $c_2=O(\delta^{-1/2})$
and hence, calculating hadronic atom spectrum at next-to-leading order in $\delta$, this
term has to be taken into account. In fact, it is nothing but the effect of the unitary
cusp in the vicinity of the elastic threshold.

In order to match the couplings $c_0,c_2$, it suffices to consider
the spin-nonflip part of the irreducible S-wave-projected
$\pi^-p$ scattering amplitude in the vicinity of the threshold, It is convenient
to perform the matching of the subthreshold amplitudes below threshold $q_0^2<0$,
where all loop diagrams $J(q_0),V_c(q_0),B_c(q_0)$ are real. In the unequal mass case, these
are given again given by Eqs.~(\ref{eq:J}), (\ref{eq:Vc}) and (\ref{eq:Bc}), respectively, with $m/2$ replaced by $\mu_c$.
The transverse photons
do not contribute at this order. Factorizing the Coulomb photon insertions in the
external lines, similar to the ones shown in Fig.~\ref{fig:Coulomb-Oalpha}a, for this
amplitude we get:
\eq
\langle\bm{p}|T^{\sf NR}_c|\bm{q}\rangle=
(1-2V_c(q_0))\left((c_0-8q_0^2c_2)+(c_0-8q_0^2c_2)^2(J(q_0)+B_c(q_0))+\cdots\right)\, .
\en
It can be straightforwardly ensured that the non-relativistic amplitude in the vicinity
of the threshold has the following behavior
\eq\label{eq:k}
(1+2V_c(q_0))\langle\bm{p}|T^{\sf NR}_c|\bm{q}\rangle
=\left(\tilde{\mathscr{A}}_c+\tilde{\mathscr{B}}_c\ln\frac{-q_0^2}{\mu_c^2}\right)
+q_0\left(\tilde{\mathscr{A}}_c'+\tilde{\mathscr{B}}_c'\ln\frac{-q_0^2}{\mu_c^2}\right)
+q_0^2\left(\tilde{\mathscr{A}}_c''+\tilde{\mathscr{B}}_c''\ln\frac{-q_0^2}{\mu_c^2}\right)+\cdots\, ,
\en
where all coefficients can be expressed in terms of $c_0,c_2$. Furthermore, according
to the matching condition, the pertinent relativistic amplitude has exactly the same behavior.
With a convenient choice of normalization, one has
\eq
\frac{1+2V_c(q_0)}{8\pi(m_p+M_\pi)}\,\langle\bm{p}|T^{\sf R}_ c|\bm{q}\rangle
=\mathscr{A}_c+\mathscr{B}_c\ln\frac{-q_0^2}{\mu_c^2}
+\cdots\, .
\en
The real
and imaginary parts of $c_0$ and $c_2$ can be extracted from the expansion of the real and imaginary parts of the irreducible $\pi^-p$ relativistic amplitude at threshold. This procedure
was discussed in detail in Ref.~\cite{Gasser:2007zt}. Here, we quote the results only.
The real part of $c_0$ at $O(\delta)$ is given by
\eq
\mbox{Re}\,c_0=\frac{2\pi}{\mu_c}\,\mbox{Re}\,\mathscr{A}_c\,\left(
1-\frac{2\pi\bar B_c}{\mu_c}\,\mbox{Re}\,\mathscr{A}_c\right)\, .
\en
The real part of $c_2$, which is of order $\delta^0$, does not contribute to the bound state energy at the accuracy we are working and will not be calculated.

Furthermore, at threshold, the elastic amplitude is complex, because there are open channels. At the order we are working, only $n\pi^0$ and $n\gamma$ channels. The ratio
of the transition cross sections in these two channels goes under the name of the
Panofski ratio and is known experimentally very accurately:
\eq
P=\frac{\sigma(\pi^-p\to\pi^0n)}{\sigma(\pi^-p\to\gamma n)}\biggr|_{\sf thr}
  \simeq 1.546\, .
  \en
  This quantity counts as $O(\delta^{-1/2})$ in NREFT, since the phase space
  in the numerator is of order $\sqrt{\Delta}=O(\delta^{1/2})$, whereas the denominator
  counts as $O(\delta)$. Note that $P$ is treated is an input in the final expression.

  The contribution of the $n\pi^0$ channel can be expressed through the spin-nonflip
  relativistic threshold
  amplitudes in the charge-exchange $\pi^-p\to\pi^0n$ and the neutral $n\pi^0\to n\pi^0$
  channels with the use of the unitarity relation. These amplitudes
  $\mathscr{A}_x,\mathscr{A}_0$ are defined by the following relations
  \eq
\frac{1+V_c(q_0)}{8\!\sqrt{2}\,\pi(m_p+M_\pi)}\,\langle\bm{p}|T^{\sf R}_x|\bm{q}\rangle
=\mathscr{A}_x+\mathscr{B}_x\ln\frac{-q_0^2}{\mu_c^2}
+\cdots\,,
\quad\quad
\frac{1}{8\pi(m_p+M_\pi)}\,\langle\bm{p}|T^{\sf R}_0|\bm{q}\rangle
=\mathscr{A}_0+\cdots\, .
\en
The matching condition for the imaginary parts gives
\eq
\mbox{Im}\,(c_0+c_0^2\bar B_c)=p^*(0)H\left(1+\frac{1}{P}\right)\, ,\quad\quad
-8\mbox{Im}\,c_2=(p^*(0))'H\, ,
\en
where
\eq
p^*(\bm{p}^2)=\frac{\lambda^{1/2}(s(\bm{p}^2),m_n^2,M_{\pi^0}^2)}{2\sqrt{s(\bm{p}^2)}}\, ,\quad\quad
s(\bm{p}^2)=(w_p(\bm{p})+w_\pi(\bm{p}))^2\, ,
\en
\eq
H=\frac{4\pi}{\mu_c}\,(\mbox{Re}\,\mathscr{A}_x)^2
\left(1+
\left(p^*(0)\mbox{Re}\,\mathscr{A}_0\right)^2\right)\, ,
\en
and $\lambda(x,y,z)=x^2+y^2+z^2-2xy-2yz-2zx$ denotes the triangle function.
It is now seen that, indeed, $\mbox{Im}\,c_2=O(\delta^{-1/2})$, since it is proportional
to the derivative of the phase space $p^*(\bm{p}^2)$.

This completes matching of all couplings of NREFT to the observables of the
underlying relativistic theory. The following remarks are due in conclusion:

\begin{itemize}
\item The $S$-matrix elements of the two-particle scattering of charged particles
  are singular at threshold in the presence of the photons, owing to the infrared
  divergences. One has to introduce a meaningful definition of the threshold amplitude by
  subtracting the divergent pieces. The definition introduced in the present section
  is a {\em convention.} One is free to use another convention. The results expressed
  in terms of the observables do not depend on a choice of the convention.
\item
  Matching can be performed in different ways, using either one-channel picture or
  coupled $\pi^-p,\pi^0n$ channels. The first case is technically easier. However, the
  counting rules in $\delta$ become more complicated, since a small mass scale $\Delta$ has
  been integrated out.
\item
  In general, the couplings in the NREFT Lagrangian are imaginary. The imaginary parts
  contain the contribution from the shielded channels and can be determined by using
  unitarity.
\item
  In principle, the irreducible amplitude is not an
  observable. We opted here to match the irreducible amplitude and the one-photon exchange amplitude separately. This is technically convenient and definitely not wrong.
  However, matching of only a sum of these two contributions would be also sufficient.
\item
  In fact, our theory is not complete because the leptons were completely ignored
  from the beginning. It turns out that the electron vacuum polarization gives a sizable
  contribution to the energy shift of the pionic hydrogen. This contribution can be however
  easily taken into account by merely modifying the Coulomb potential between
  the proton and the charged pion (the so-called Uehling potential). There is no need
  to include electrons explicitly in the Lagrangian.
  
\end{itemize}

\subsection{Bound state equation and the spectrum}

The Coulomb photons are not dynamical degrees of freedom, and can be integrated
out, leading to a non-local Lagrangian. Using canonical formalism, one can derive
the Hamiltonian which consists of the Coulomb interaction between charged particles
and the rest:
\eq
   {\bf H}={\bf H}_0+{\bf H}_C+{\bf H}_I\, ,\quad\quad
   {\bf H}_A(t)=\int d^3\bm{x}\,\mathscr{H}_A(\bm{x},t)\, ,\quad A=0,C,I\, .
   \en
   Here, $\mathscr{H}_0$ is the free Hamiltonian of protons, negatively charged
   pions and transverse photons. The Coulomb potential is given by a non-local expression
   \eq\label{eq:C}
   \mathscr{H}_C=4\pi\alpha(\pi_-^\dagger\pi_-)\Delta^{-1}(\psi^\dagger\psi)\, .
   \en
   and the Hamiltonian $\mathscr{H}_I=\mathscr{H}_\gamma+\mathscr{H}_R+\mathscr{H}_S+\mathscr{H}_{\sf vac}$ consists of several pieces:
   \eq
   \mathscr{H}_\gamma&=&-\frac{ie}{2m_p}\,\psi^\dagger(\nabla\bm{A}+\bm{A}\nabla)\psi
   +\frac{ie}{2M_\pi}\,\pi_-^\dagger(\nabla\bm{A}+\bm{A}\nabla)\pi_-\, ,
   \nonumber\\[2mm]
   \mathscr{H}_R&=&
   -\psi^\dagger\frac{\nabla^4}{2m_p}\,\psi-\pi_-^\dagger\frac{\nabla^4}{2M_\pi}\,\pi_-\, ,
   \nonumber\\[2mm]
   \mathscr{H}_S&=&-\tilde c_0(\psi^\dagger\pi_-^\dagger)(\pi_-\psi)
-c_2\left((\psi^\dagger\lrarrowD^2\pi_-^\dagger)(\pi_-\psi)+\mbox{h.c.}\right)\, ,
   \en
   where $\tilde c_0=c_0-e^2\lambda$, and
   \eq
   \lambda=\frac{c_p^D}{8m_p^2}+\frac{h_1}{6M_\pi^2}=\frac{1+2\mu_p}{8m_p^2}
   +\frac{1}{6}\,(\langle r_p^2\rangle+\langle r_\pi^2\rangle)\, .
   \en
   The term proportional to $\lambda$ emerges, when Coulomb photons are integrated out. Furthermore, $\mathscr{H}_{\sf vac}$ stands for the electron vacuum polarization contribution,
   which stems from the modification of the Coulomb potential in $\mathscr{H}_C$ given
   in Eq.~(\ref{eq:C}):
   \eq
   -\frac{\alpha}{r}\to-\frac{\alpha}{r}-\frac{4 \alpha^2}{3}\,
   \int\frac{d^3\bm{k}}{(2\pi)^3}\,e^{i\bm{k}\bm{r}}
   \int_{4m_e^2}^\infty\frac{ds}{s+\bm{k}^2}\frac{1}{s}\left(1+\frac{2m_e^2}{s}\right)
   \sqrt{1-\frac{4m_e^2}{s}}\doteq V_C(r)+V_{\sf vac}(r)\, .
   \en
Here, $m_e$ is the electron mass.

Calculation of the energy spectrum of hadronic atoms is then straightforward. The Hamiltonian ${\bf H}_0+{\bf H}_C$ produces the pure Coulomb spectrum in the $p\pi^-$ sector
plus any number of non-interacting transverse photons. The ground-state
eigenvector, corresponding to the pionic hydrogen, in the CM frame is given by
\eq
|\Psi_0,\sigma\rangle = \int \frac{d^3\bm{p}}{(2\pi)^3}\,\Psi_0(\bm{p})
b^\dagger(\bm{p},\sigma)a^\dagger(-\bm{p})|0\rangle\, ,\quad\quad
\Psi_0(\bm{p})=\frac{(64\pi\gamma^5)^{1/2}}{(\bm{p}^2+\gamma^2)^2}\, ,\quad
\gamma=\alpha\mu_c\, .
\en
Here, $b^\dagger(\bm{p},s)$ and $a^\dagger(-\bm{p})$ denote the creation operators
for the non-relativistic proton and $\pi^-$, respectively, $\sigma$ is the projection
of the proton spin on the third axis, and $\Psi_0(\bm{p})$ stands for the
pure Coulomb ground-state wave function. The pure Coulomb Green function
has a pole at the bound state energy $E_1=m_p+M_\pi-\mu_c\alpha^2/2$, see
Eq.~(\ref{eq:hydrogen}). This pole
is shifted, when the interaction ${\bf H}_I$ is turned on. The shift can be calculated
by using standard Rayleigh-Schr\"odinger perturbation theory in quantum mechanics.
A compact equation to determine the shifted pole position $E$ in the complex plane
is given by
\eq\label{eq:master}
E-E_1-\langle \Psi_0|{\bf U}(E)|\Psi_0\rangle=0\, ,\quad\quad
\langle \Psi_0|{\bf U}(E)|\Psi_0\rangle=\frac{1}{2}\,\sum_\sigma
\langle \Psi_0,\sigma|{\bf U}(E)|\Psi_0,\sigma\rangle\, .
\en
Here, the operator ${\bf U}(E)$ obeys the equation
\eq\label{eq:U}
   {\bf U}(E)={\bf H}_I+{\bf H}_I\hat {\bf G}_C(E){\bf U}(E)\, ,
   \quad\quad
   \hat {\bf G}_C(E)=\frac{1}{E-{\bf H}_0-{\bf H}_C}-\sum_\sigma\frac{|\Psi_0,\sigma\rangle\langle \Psi_0,\sigma|}{E-E_1}\, .
   \en
  The operator   $\hat {\bf G}_C(E)$ denotes the ``pole-removed'' Coulomb Green function. The
   presence of this function guarantees that the sums over the intermediate states in the
   usual quantum mechanical perturbation series do not contain the contribution from the unperturbed ground state.

   The calculations proceed as follows: one iterates Eq.~(\ref{eq:U}) in order to obtain
   an expansion of ${\bf U}(E)$ in terms of ${\bf H}_I$.
   As in the scattering sector, threshold expansion should be applied to all expressions,
   appearing in this expansion. Then, power counting is restored and the subsequent terms
   contribute to higher orders in $\delta$. Hence, at a given order in $\delta$,
   this expansion can be truncated. At the next step, one expands the matrix element in
   Eq.~(\ref{eq:master}) in $(E-E_1)$ and finds an explicit expression of this quantity
   at the order in $\delta$ one is working. These (pretty standard) calculations have been
   described at many places (see, e.g., the review paper~\cite{Gasser:2007zt} and references therein) and will not be repeated here. The final result at the next-to-leading
   order in $\delta$ is rather compact and is, by convention, split into the ``electromagnetic'' and ``strong'' parts:
   \eq\label{eq:DeltaE}
   (E-E_1)=\left[-\frac{5}{8}\,\alpha^4\mu_c^4
   \frac{m_p^3+M_\pi^3}{m_p^3M_\pi^3}-\frac{\alpha^4\mu_c^3}{m_pM_\pi}
   +4\alpha^4\mu_c^3\lambda+\Delta_{\sf vac}\right]_{\sf em}
 +\left[-\frac{\alpha^3\mu_c^3}{\pi}\,\left(c_0(1+\delta_{\sf vac})+8c_2\gamma^2
     +c_0^2\left(\bar B_c-\frac{\alpha\mu_c^2}{\pi}\,(\ln\alpha-1)\right)\right)\right]_{\sf str}\, .
   \en
   Here,
   \eq
   \Delta_{\sf vac}=\int d^3\bm{r}\,\Psi_0^2(\bm{r})V_{\sf vac}(r)\simeq -3.241\,\mbox{eV}\, ,\quad\quad
   \delta_{\sf vac}=2\int d^3\bm{r}\,\Psi_0(\bm{r})\langle \bm{r}|\hat{\bf G}_C(E_1)|\bm{0}\rangle\simeq 4.8\cdot 10^{-3}
   \en
   are the corrections arising from the vacuum polarization
   which can be evaluated, given the explicit expression of $V_{\sf vac}(r)$ (see, e.g., Ref.~\cite{Eiras:2000rh}). In the counting, they come
   at the same order in $\delta$ as other corrections, if one counts
   $m_e/M_\pi$ as $O(\delta)$.
   Note also that the spin-averaged Coulomb Green function in the coordinate space
   obeys the relation $\langle\bm{0}|\hat{\bf G}_C(E_1)|\bm{0}\rangle=-\bar B_c+\alpha\mu_c^2(\ln\alpha-1)/\pi$ at the origin. This relation was used for
   derivation of Eq.~(\ref{eq:DeltaE}).

   What remains is to express all couplings in Eq.~(\ref{eq:DeltaE}) in terms of the
   physical observables through the matching condition. The final result is remarkably
   simple. The real and imaginary parts of $(E-E_1)_{\sf str}$, which determine the strong
   energy shift of the bound state and its width, are given by
   \eq
   (E-E_1)_{\sf str}=\Delta E_{\sf str}-\frac{i}{2}\,\Gamma
   =-2\alpha^3\mu_c^2\mathscr{A}_c(1-2\alpha\mu_c(\ln\alpha-1)\mathscr{A}_c+\delta_{\sf vac})\, .
   \en
   Here, in order to obtain an explicit formula for the width, one has to use here the
   expression for $\mbox{Im}\,\mathscr{A}_c$, obtained earlier from unitarity. One finally gets:
   \eq\label{eq:Deser-final}
   \Delta E_{\sf str}&=&-2\alpha^3\mu_c^2\mbox{Re}\,\mathscr{A}_c
   (1-2\alpha\mu_c(\ln\alpha-1)\mbox{Re}\,\mathscr{A}_c+\delta_{\sf vac})\, ,
   \nonumber\\[2mm]
   \Gamma&=&8\alpha^3\mu_c^2
p^*(-\gamma^2)
   (\mbox{Re}\,\mathscr{A}_x)^2
\left(1-4\alpha\mu_c(\ln\alpha-1)\mbox{Re}\,\mathscr{A}_c+
\left(p^*(0)\mbox{Re}\,\mathscr{A}_0\right)^2+\delta_{\sf vac}\right)\left(1+\frac{1}{P}\right)\, .
\en
These are the final expressions for the energy shift and the width of the ground state of
pionic hydrogen at next-to-leading order in the isospin breaking parameter $\delta$,
which is obtained in NREFT.

\subsection{Isospin-breaking corrections in QCD+QED}

The threshold amplitudes $\mathscr{A}_c,\mathscr{A}_x,\mathscr{A}_0$ at leading
order in isospin breaking are expressed through linear combinations of the S-wave $\pi N$ scattering lengths. At order $\delta$, these relations take the following form (we use the notations of Ref.~\cite{Schopper:1983hnv}):
\eq
\mathscr{A}_c=a_{0+}^++a_{0+}^-+\delta_c+\cdots\, ,
\quad\quad
\mathscr{A}_x=-a_{0+}^-+\delta_x+\cdots\, ,
\quad\quad
\mathscr{A}_0=a_{0+}^++\delta_0+\cdots\, .
\en
Here, ellipses stand for the higher-order terms in $\delta$. Our final aim is to extract the
pion-nucleon scattering lengths $a_{0+}^\pm$ from the measured values of the
energy shift and width.

In ChPT with virtual photons it is possible to evaluate $\delta_c,\delta_x,\delta_0$ at the first
order in $\delta$ and order by order in the expansion of quark masses (in fact, at the accuracy
we are working, $\delta_0$ is not needed). Given a definite prescription for splitting the
strong and electromagnetic interactions, all these quantities are uniquely defined.
For example, in Ref.~\cite{Gasser:2002am} calculations were carried out at $O(p^3)$ in ChPT.
The result for the energy shift is written in the following form:
\eq
\Delta E_{\sf str}=-2\alpha^3\mu_c^2(a_{0+}^++a_{0+}^-)(1+\delta_\epsilon)\, ,\quad\quad
\delta_\epsilon=(-7.2\pm 2.9)\cdot 10^{-2}\, .
\en
Note a sizable error in the correction term, which is mainly caused by a poor knowledge of
some low-energy constants. A huge advantage of the NREFT approach consists in the fact that
this uncertainty does not emerge at leading order in $\delta$, where the result to all
orders is written down in terms of the scattering lengths.
Furthermore, the first-order corrections do the decay width were calculated
in~\cite{Zemp:2004zz}. Isospin-breaking corrections to the amplitude
in all physical channels were evaluated, e.g., in Refs.~\cite{Hoferichter:2009gn,Hoferichter:2009ez}.
Using, in particular, these results,
both pion-nucleon scattering lengths were extracted from a combined analysis of the
pionic hydrogen energy shift and the pionic deuterium energy shift~\cite{Baru:2011bw,RuizdeElvira:2017stg,Hoferichter:2015hva,Hoferichter:2016ocj,Hoferichter:2023ptl}.

Similar path can be pursued for different hadronic atoms. In some cases, a slight modification
will be needed. Namely, if the threshold of a neutral channel lies above the charged one
(like, e.g., the kaonic hydrogen, which cannot decay into a neutron and a neutral kaon),
a unitary cusp shows up in the real part of the amplitude. The leading-order isospin-breaking
corrections in this case start at $O(\sqrt{\delta})$ and not at $O(\delta)$. These corrections
are however of a purely kinematic origin and can be expressed in terms of the scattering
lengths in a model-independent way~\cite{Meissner:2004jr}.

\section{Conclusions}

\begin{itemize}

\item
  Experiments on hadronic atoms allow one to extract precise values of the strong
  scattering lengths from the measured values of the energy level shift and width.
  This possibility emerges due to the existence of two disparate distance scales:
  in an atom that is bound mainly by a Coulomb force, the hadrons are located
  very far of each other (at a hadronic scale) and, at such distances, their
  interaction, at a very good approximation, can be described in terms of scattering
  lengths only.
  In order to be able to perform this beautiful test, however, one should be able to
  carry out the theoretical calculation of these observables in the accuracy that matches
  experimental precision. In particular, isospin-breaking corrections of order $\alpha$
  and $m_d-m_u$ are relevant, and should be evaluated in the theory that one is going
  to test (i.e., QCD). The results obtained in this way are devoid of model-dependence
  and yield a reliable estimate for the theoretical uncertainty.

\item
  ChPT with virtual photons is a low-energy effective theory of QCD plus QED and thus describes (in principle) the predominately electromagnetic bound states of hadrons very close
  to the elastic threshold. However, the bound-state problem, which is non-perturbative
  by definition, is notoriously complicated in ChPT, as in any relativistic field theory. The
  NREFT, which is an effective theory of ChPT at non-relativistic momenta (much smaller
  than the mass of a lightest massive particle in the system) is perfectly suited to do this
  job. Namely, on one hand, the poles corresponding to hadronic atoms, lie in the region
  of validity of NREFT. On the other hand, description of the bound states in NREFT is much
  simpler. In fact, the usual quantum-mechanical Rayleigh-Schr\"odinger perturbation
  theory can be applied, with subsequent terms corresponding to higher orders in the
  isospin-breaking parameter $\delta\sim\alpha,(m_d-m_u)$. The expansion can be
  truncated after the required precision is reached.

\item
  The calculations proceed as follows. At the first stage, we perform matching of the NREFT
  couplings to the threshold $S$-matrix elements in the relativistic theory. It is very
  important to note that the matching can be done perturbatively. Only a finite number
  of diagrams in NREFT give nonzero contribution to the threshold scattering amplitude
  and, hence, the obtained matching condition is exact to all orders. Furthermore,
  in the presence of photons, the $S$-matrix elements are infrared-divergent. The
  matching condition is written down in terms of the finite parts thereof. At the next step,
  the spectrum of the bound states is calculated in NREFT by using standard
  Rayleigh-Schr\"odinger perturbation theory. Then, with the use of matching, the effective
  NREFT couplings are replaced by relativistic threshold amplitudes, and any trace
  of the NREFT disappears in the final expressions. At the final stage, one uses ChPT
  to evaluate isospin-breaking corrections in the threshold amplitudes in a systematic expansion in quark masses.

\item
  The potential model has been extensively used in the past to calculate isospin-breaking
  corrections to the lowest-order DGBT formula. To this end, usually,
  a ``realistic,'' isospin-symmetric
  hadronic potential was amended by a Coulomb interaction, physical masses
  were assigned to the particles, and the corrections to the spectrum have been
  calculated. There is a conceptual difference between this method and the one we are
  using. Namely, the isospin breaking in ChPT includes {\em all} sources of breaking in
  QCD, parametrizing short-range effects from virtual photon exchange and $(m_d-m_u)$
  in terms of certain low energy constants. Even some of them might be poorly known,
  this approach enables one to fairly quantify systematic uncertainty in the final result.
  On the contrary, potential approach is a model, where short-range isospin-breaking
  effects are typically absent at all. Since these are, as a rule, of the same order of magnitude as the effects that are included, one may conclude that there is an inherent systematic uncertainty present in the potential model, which is very hard to control, see Ref.~\cite{Lipartia:2001zh} for a more detailed discussion.

\item
  The systems that contain two disparate scales are very interesting, both experimental
  and theoretically. In the introduction, we have mentioned several such systems in different fields of physics. Here we want to stress that the NREFT methods, used to deal with
  hadronic atoms, are universal and can be extended, without much modification, to other systems as well.

\end{itemize}

\section{A brief guide to the literature}

Taking into account the format of the present work, it is impossible to give credit
to all work done to create a theory of hadronic atoms.
We therefore restrict ourselves to only few
important papers that have marked significant development of the concepts and methods.

As mentioned above, originally the DGBT formula was derived in the potential scattering
theory~\cite{Deser:1954vq}. The relation between the scattering length and the
bound-state energy of a hadronic atom, described by this formula, is universal and is
determined only by the fact that the electromagnetic and hadronic scales are very
distinct. Trueman~\cite{Trueman:1961zza} was the first to apply potential model
to the study of electromagnetic corrections. He obtained a regular expansion of the
energy shift in powers of $R/r_B$, where $R$ stands for a typical range of strong
interactions. In particular, his result correctly reproduces the $\ln\alpha$ term in the
energy shift in Eq.~(\ref{eq:Deser-final}). The references given here include
important contributions to the field~\cite{Deloff:2003ns,Batty:1997zp,Oades:2007de,Gashi:2001wv,Gashi:1997ck,Ericson:2004ps}.

As already mentioned, the potential method
does not include short-range isospin-breaking contributions and thus possesses
a built-in systematic uncertainty which is difficult to control. After realizing this fact,
attempts were made to describe the spectrum of hadronic atoms in low-energy effective
theories of QCD~\cite{Krewald:2003ab,Jallouli:2006an,Jallouli:1997ux,Ivanov:1998wx,Lyubovitskij:1996mb,Hammer:1999up}. Albeit these developments were a step forward towards a systematic inclusion of the isospin-breaking effects, carrying out calculations in
the relativistic bound-state framework was a rather challenging enterprise.
Finally, it has been realized that the non-relativistic EFT provides an natural
framework for the description of this sort of hadronic bound
systems~\cite{Kong:1998xp,Kong:1999wm,Holstein:1999nq}.
A complete expression of the pionium decay width was obtained in this approach~\cite{Gall:1999bn}, and an accurate numerical analysis has been carried out~\cite{Gasser:1999vf}.
The few following years have seen an essential surge of activity in the field that
laid foundation for the theory of hadronic atoms as known at present~\cite{Gasser:2001un,Eiras:2000rh,Gasser:2002am,Meissner:2004jr,Labelle:1998gh,Eiras:1999xx,Baru:2011bw,Zemp:2004zz,Meissner:2006gx,Meissner:2005bz,Schweizer:2004qe,Hoferichter:2009gn,Hoferichter:2009ez,RuizdeElvira:2017stg,Hoferichter:2015hva,Hoferichter:2016ocj,Hoferichter:2023ptl}. The review articles~\cite{Gasser:2007zt,Gasser:2009wf} summarize these developments and contain  a detailed bibliography
on the subject.

          \begin{ack}[Acknowledgments]%

            The author would like to thank Jürg Gasser and Ulf-G. Mei{\ss}ner for careful reading of the manuscript and useful suggestions. Furthermore,           
the author acknowledges financial support
from the Ministry of Culture and Science of North Rhine-Westphalia through the
NRW-FAIR project  and from the Chinese Academy of Sciences (CAS) President's
International Fellowship Initiative (PIFI) (grant no. 2024VMB0001).

\end{ack}



\bibliographystyle{Numbered-Style} 
\bibliography{myreference}

\end{document}